# AN EXTENDED STABLE MARRIAGE PROBLEM ALGORITHM FOR CLONE DETECTION

Hosam AlHakami, Feng Chen and Helge Janicke

Software Technology Research Laboratory, De Montfort University, Leicester, UK
hosam.alhakami@myemail.dmu.ac.uk
{fengchen,heljanic}@dmu.ac.uk

## ABSTRACT

*Code cloning negatively affects industrial software and threatens intellectual property. This paper presents a novel approach to detecting cloned software by using a bijective matching technique. The proposed approach focuses on increasing the range of similarity measures and thus enhancing the precision of the detection. This is achieved by extending a well-known stable-marriage problem (SMP) and demonstrating how matches between code fragments of different files can be expressed. A prototype of the proposed approach is provided using a proper scenario, which shows a noticeable improvement in several features of clone detection such as scalability and accuracy.*

## 1. INTRODUCTION

The Stable Marriage Problem (SMP) is a well-known problem that has been defined by Gale and Shapley in 1962 [1]. An example of the SMP is allocating the right jobs to their most suitable jobseekers (one-one). Similarly framed problems with differing cardinality are also considered to be instances of the SMP, such as matching graduated medical students to hospitals (one-many) [2]. The SMP grantees the stable match between the candidates.

Clone detection has been intensively investigated due to the need of tackling code issues in the maintenance process. Current detection algorithms are search-based algorithms that do not consider finding a match between a given code segment in a larger set of code files. Our approach differs in that it uses the preferences of candidates (code portions) in the process of finding the best matches.

In this paper, a variant of the stable marriage problem algorithm to clone detection is investigated to find clones of different source files. The extended algorithm introduces the preferences of code segments based on the values of predefined metrics, e.g. the number of calls from or to a method, cyclomatic complexity. The values of both parties will be considered in the clone detection process.

The remainder of this paper is structured as follows. In Section 2, the background of the SMP research is introduced. In Section 3, the context for the SMP algorithms to be applied to Clone Detection is discussed. In Section 4, an adapted SMP algorithm that is suitable to generate fair and stable matches between similar code fragments of different source files is proposed and evaluated, and we conclude the paper in Section 5.

## 2. SMP Algorithm

In 1962, David Gale and Lloyd Shapley published their paper College admissions and the stability of marriage [1]. This paper was the first to formally define the Stable Marriage Problem (SMP), and provide an algorithm for its solution. The SMP is a mechanism that is used to match two sets of the same size, considering preference lists in which each element expresses its preference over the participants of the element in the opposite set [1]. Thus, the output has to be stable, which means that the matched pair is satisfied and both candidates have no incentive to disconnect. A matching $M$ in the original SMP algorithm is a one-to-one correspondence between the men and women. If man $m$ and woman $w$ are matched in $M$, then $m$ and $w$ are called partner in $M$, and written as $m = PM(w)$ (which is the M-partner of $w$), $w = PM(m)$ (the M-partner of $m$). A man $m$ and a woman $w$ are said to block a matching $M$, or called a blocking pairs for $M$ if $m$ and $w$ are not partners in $M$, but $m$ prefers $w$ to $PM(m)$ and $w$ prefers $m$ to $PM(w)$ [2]. Therefore, a matching $M$ is stable when all participants have acceptable partners and there is no possibility of forming blocking pairs. This problem is in interest of a lot of researchers in many different areas from several aspects. Matching problems on bipartite sets where the entities on one side may have different sizes are intimately related to the scheduling problems with processing set restrictions [3].

An instance $I$ of SM involves $n$ men and $m$ women, each of whom ranks all $n$ members of the opposite sex in strict order of preference. In $I$ we denote the set of men by $m = m_1, m_2, ..., m_n$ and the set of women by $w = w_1, w_2, ..., w_n$. In SM the preference lists are said to be complete, that is each member of $I$ ranks every member of the opposite sex as depicted in figure 1.

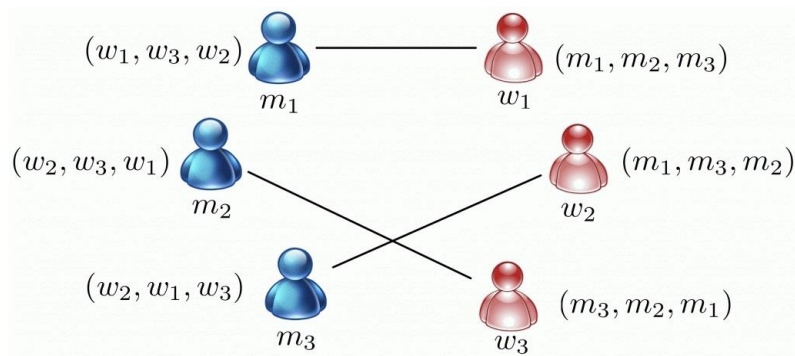

Figure1. General view of SMP. [4]

### 2.1. Gale Shapley Extended Algorithm

The algorithm presented by Gale and Shapley for finding a stable matching uses a simple deferred acceptance strategy, comprising proposals and rejections. There are two possible orientations, depending on who makes the proposals, namely the man-oriented algorithm and the woman-oriented algorithm.

In the man-oriented algorithm, each man $m$ proposes in turn to the first woman $w$ on his list to whom he has not previously proposed. If $w$ is free, then she becomes engaged to $m$. Otherwise, if $w$ prefers $m$ to her current fiancé $m$, she rejects $m$, who becomes free, and $w$ becomes engaged to $m$. Otherwise $w$ prefers her current fiancé to $m$, in which case $w$ rejects $m$, and $m$ remains free. This process is repeated while some man remains free. For the woman-oriented algorithm the process is similar, only here the proposals are made by the women.

The man-oriented and woman-oriented algorithms return the man-optimal and woman-optimal stable matching respectively. The man-optimal stable matching has the property that each man obtains his best possible partner in any stable matching. However, while each man obtains his best possible partner, each woman might simultaneously obtain her worst possible partner in any stable matching. Correspondingly, the woman-oriented algorithm has the same problem.

**Theorem 1** All possible execution of the Gale-Shapley algorithm (with the men as proposers) yields the same stable matching, and in this stable matching, each man has the best partner that he can have in any stable matching [2].

According to the previous theorem if each man has given his best stable partner, then the result is a stable matching. The stable matching generated by the man-oriented version of the Gale-Shapely algorithm is called man-optimal. However, in the man-optimal stable matching, each woman might have the worst partner that she can have in any stable matching, leading to the terms of man-optimal is also woman-pessimal. This results in the next theorem.

**Theorem 2** In the man-optimal stable matching, each woman might have the worst partner that she can have in any stable matching [2].

The following example in Figure 2 gives the different output for both man-optimal and woman-optimal, the instance formed out of 4 elements.

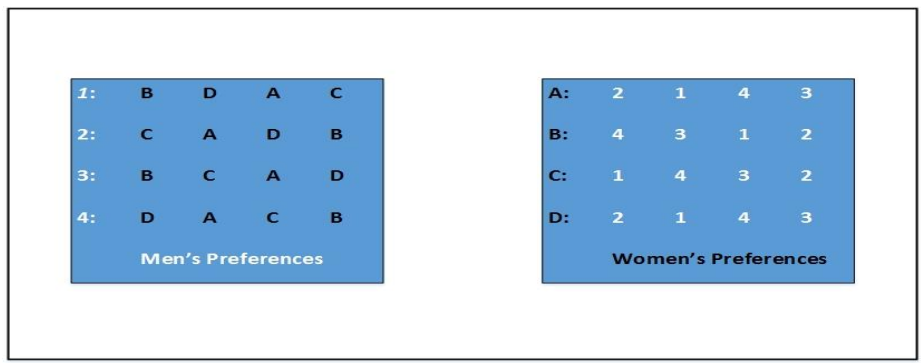

Figure 2. A stable marriage instance of size 4.

The results of different cases differ from man-oriented version to woman-oriented version. The stable matching generated by both man-oriented and women-oriented versions are respectively
Man-Optimal = M0 = {(1 , D) , (2 , C) , (3 , B) , (4 , A)} and
Woman-optimal= Mz = {(1 , D) , (2 , A) , (3 , B) , (4 , C)}.

**Algorithm 1** Extended Gale-Shapley algorithm [2]
1: assign each person to be free
2: **while** some man $m$ is free **do**
3: begin
4: $w$:= first woman on $m$'s list;
5: **if** some man $p$ is engaged to $w$ **then**
6: assign $p$ to be free;
7: assign $m$ and $w$ to be engaged to each other;
8: **for each** successor $m'$ of $m$ on $w'$ list **do**
9: delete the pair($m'$,$w$)
10: end;

An extended version of Gale-Shapley algorithm has been designed to improve the basic algorithm. The extended version reduces the preference list by eliminating specific pairs that can be clearly identified as unrelated to any stable matching. The deletion process of such pair is performed by deleting each other from the preference lists.

### 2.2. HOSPITALS/RESIDENTS PROBLEM

The hospitals/residents problem (also called Colleges/Students problem, and by many other names) reflects a cardinality of many-to-one of the stable marriage problem. This cardinality touches a wide range of large-scale applications that require stable matching such as students/colleges problem. Therefore, it has interested the researchers in different aspects for instance recruitment in which uses schemes to match a group and employers to a group of employees. The National Resident Match Program [5] is a real example in the US which annually matches hospitals to about 30,000 medical residents. An instance of the hospitals/residents (HR) problem consists of a set $R$ of $n$ residents and a set $H$ of $m$ hospitals, where each hospital $h$ has capacity $ch$, the maximum number of positions available in $h$. Each resident ranks the hospitals in $H$ that are acceptable to him/her in strict order of preference; likewise, each hospital ranks the residents in $R$ that are acceptable to it in strict order of preference. A matching $M$ for the instance is a set of resident-hospital pairs where in every pair the resident and the hospital are mutually acceptable to each other, every resident appears in at most one pair, and every hospital $h$ appears in at most $ch$ $(r, h) \notin M$ pairs. A pair forms a blocking pair with respect to $M$ if

i) $r$ is unmatched and finds $h$ acceptable or $r$ prefers $h$ to the hospital she is assigned to and, simultaneously,
ii) $h$ is not filled to capacity and finds $r$ acceptable or $h$ prefers $r$ to one of the residents assigned to it.

---
**Algorithm 2** Hospital-oriented algorithm
1: assign each resident to be free
2: assign each hospital to be totally unsubsidised;
3: **while** (some hospital $h$ is unsubsidised) and ($h$'s list contains a resident $r$ not provisionally assigned to $h$) **do**
4: begin
5: $r$:= first such resident on $h$'s list;
6: **if** $r$ is already assigned, say to $h$', **then**
7: break the provisional assignment of $r$ to $h$';
8: provisionally assign $r$ to $h$;
9: **for each** successor $h$' of $h$ on $r$'s list **do**
10: remove $h$' and $r$ from each other's lists
11: end;

---

Intuitively, if *(r, h)* forms a blocking pair in *M* then *r* and *h* are likely to break their assignments under *M*, causing the matching to unravel. Thus, the goal of the HR problem is to find a matching that is stable and has no blocking pairs. In their seminal paper [1], Gale and Shapley first tackled the problem in the simpler stable marriage (SM) setting where residents and hospitals are replaced by men and women. Every participant has a complete preference list (i.e., every man ranks all the women and every woman ranks all the men), and a capacity of one (i.e., every individual can have at most one assigned partner). They introduced the deferred-acceptance algorithm to find a stable matching, and showed that the algorithm can be extended to the more general HR setting. Consequently, they proved that every HR instance has a stable matching which can be computed in *O (nm)* time. In [6] Cheng et al. examined the structure of the set of all stable matchings of an HR instance and introduce the notion of meta-rotations in this setting. Also, they discuss the problem of finding feasible stable matchings.

**Theorem 3** ([2])
(i) The matching specified by the provisional assignments after the execution of the hospital-oriented algorithm is stable.
(ii) In this matching, a hospital $h$ with $q$ available places is assigned either its best $q$ stable partners, or a set of fewer than $q$ residents; in the latter case no other resident is assigned to $h$ in any stable matching.
(iii) Each resident is assigned in this matching to his worst stable partner.

We will build on this background in Section 4 where we use the extended Gale-Shapley algorithm in our application to the problem of clone detection.

## 3. CLONE DETECTION

Clone detection is a crucial field that has been intensively conducted by researchers and practitioners for the last two decades to enhance a software systems work and therefore, improves the maintainability for the future lifespan of the software system. Although the clone detection is a wide spread research problem over many years, it is still considered as a fuzzy terminology since the researchers have differently defined it according to variants situations and criteria. Thus, it is essential to understand the meaning and usage of the clones to know how to deal with it properly. In this section, we provide different definitions and types of clones.

### 3.1. CLONE RELATION TERMS

Clone is usually detected as a form of either clone pair or clone classes. These two terms focus on the similarity relation between two or more pieces of cloned code. Kamiya et al. in [7] describe this relation as an equivalence relation (i.e., a reflexive, transitive, and symmetric relation). It can be said that there is a clone-relation between two fragments of code if (and only if) they have the same sequences (original characters strings, strings without whitespaces, token type etc.). We can express the meaning of clone pair and clone classes based on the clone relation as shown in Figure 3 below:

| Fragment 1: | | Fragment 2: | | Fragment 3: | |
|---|---|---|---|---|---|
| ... | | ... | | ... | |
| for (int i=1; i<n; i++) {<br>  sum = sum + i;<br>} | a | for (int i=1; i<n; i++) {<br>  sum = sum + i;<br>} | a | | |
| if (sum <0 ) {<br>  sum = n - sum;<br>} | b | if (sum <0 ) {<br>  sum = n - sum;<br>} | b | if (result <0 ) {<br>  result = m - result;<br>} | a |
| ... | | while ( sum < n ) {<br>  sum = n / sum ;<br>} | c | while (result < m ) {<br>  result = m / result<br>} | b |
| | | ... | | ... | |

Figure 3. Clone pair and Clone class. [8]

- **Clone Pair:** two fragments of code are considered to form a clone pair when they have a clone-relation between them. That means these two portions are either identical or similar to each other. As seen in Figure 3 for the three code fragments, Fragment 1 *(F1)*, Fragment 2 *(F2)* and Fragment 3 *(F3)*, we can get five clone pairs, *(F1(a), F2(a))*, *(F1(b), F2(b)), (F2(b), F3(a)), (F2(c), F3(b))* and *(F1(b), F3(a))*. If we assume to extend the granularity size of cloned fragments, we get basically two clone pairs, *(F1(a + b), F2(a + b))* and *(F2(b + c), F3(a + b))*. And if we consider the granularity not to be fixed, we get seven clone pairs, *(F1(a), F2(a)), (F1(b), F2(b)), (F2(b), F3(a)), (F2(c), F3(b)), (F1(b), F3(a)), (F1(a+b), F2(a+b))* and *(F2(b + c), F3(a + b))*; each of these fragments is termed as a simple clone [9].
- **Clone Class:** is a maximal set of related portions of code that contains a clone pairs. It can be seen that the three code fragments of Figure 3, we get a clone class of *(F1(b), F2(b), F3(a))* where the three code portions *F1(b), F2(b)* and *F3(a)* form clone pairs with each other *(F1(b), F2(b)), (F2(b), F3(a))* and *(F1(b), F3(a))* result in three clone pairs. Consequently, a clone class is the union of all clone pairs which have portions of code in common [10, 11].
- **Clone Communities:** as termed in [12], it is another name of the Clone classes that reflecting the aggregation of related code fragments which form a clone pairs.
- **Clone Class Family:** researchers in [10] revealed the term of clone class family to group or aggregation of all clone classes that have the same domain.
- **Super Clone:** as have been outlined by [13] multiple clone classes between the same source entities (subsystems or clone classes) are aggregated into one large super clone which is the same as the clone class family.
- **Structural Clones:** it is an aggregation of similar simple clones that spread in different clone classes in the whole system [9]. Therefore, it can be classified as both a class clone (in early stage of clustering similar fragments of code) and super clone.

## 3.2. DEFINITION OF CODE CLONING

As aforementioned there is no original or specific definition of cloned code and therefore, all anticipated clone detection methods have their own definition for code clone [14, 15]. However, a fragments of code that has identical or similar code fragments in the source code, is considered to be a code clone. Regardless the changes that have been applied on a certain code clone, if still in the thresholds of the copied portion, then both the original and the copied fragments term as code clones and they form a clone pair.

Some researchers based their definition of clone code on some definition of Similarity whereas there is no specified definition of detection independent clone similarity. Baxter, Yahin et al. [16] defined code clones as the fragments of code that are similar based on definition of similarity and they provide a threshold-based definition of tree similarity for near-miss clones. However, there is a fuzziness of the term similarity; what is meant by similar? , and to what extend are they similar? The definition provided by Kamiya et al. [7] zooms in this terminology as they define the clones as the segments of source files that are identical or similar to each other. Another ambiguous definition is proposed by Burd and Cordy [11, 17] in which fragment of code called clone when there is more existences of that fragment in the source code with or without minor modifications. However, a number of researchers Kontogiannis, Lu et al. and Kapser in [15, 18, 19] tried to control and specify their own detection dependent threshold based definition of the term similarity. Therefore, after several comparisons that run-out by Roy, Kontogiannis and Koschke et al. [11, 15, 20] they attempt to automatically unify the result sets of multiple detectors, trying to solve the differential detector-based output.

## 3.3 CLONE DETECTION TECHNIQUES AND TOOLS

The following are some features (dimensions) which form corner stones and clarify the several facets of the clone detection techniques:

- **Source Representation:** it is the final form or representation of the code fragments being compared in the comparison phase to meet the algorithm requirements. This can be achieved using the different types of transformations/normalizations or filtering.

- **Source Transformation/Normalization:** some clone detection approaches apply a kind of transformation / normalization or filtering on the original source code to get a suitable code format in order to apply a comparison algorithm. However, some other approaches only remove the comments and the whitespaces.
- **Comparison Algorithm:** one of the major concern properties that can affect the performance of the clone detection process is the choice of the appropriate detection or matched algorithm. Several approaches apply different data mining/information retrieval algorithms.
- **Clone Granularity:** there are two types of granularity clones fixed and free granularity. A returned fixed granularity clone is the pre-defined block size of the considered code fragments such as (e.g., function, begin-end brackets etc.). However, if there is no certain considered limit or size of the code portion block then they are called free granularity clones.
- **Clone Similarity:** This feature represents the type of code clones that can be found by some detection techniques such as exact match clone, parameterized match or near-miss clones.

### 3.3.1 TEXT-BASED TECHNIQUE

This technique is purely based on the text or string methods, so in this approach the raw source code is considered as sequence of lines and strings. Code segments are matched with each other to detect the same sequences of text or strings, which not related to structural elements of the language. The detected sequences are returned as clone pair by the detection technique. Some text-based approaches perform a slight of transformation/normalization on the code fragments before setting off the comparison process, whereas normally the row source code is directly used in the matched process. The following are some commonly used filtering/transformation/normalization in some approaches:

- **Normalization**: basic normalization can be applied on the raw source code (see Figure4).
- **Whitespaces**: considers and removes all whitespaces including tabs, new line(s) and other blanks spaces.
- **Comments**: eliminates all comments used in the source code.

| Operation | Language element | Example | Replacement |
|---|---|---|---|
| 1 | Literal string | "Abort" | "..." |
| 2 | Literal character | 'y' | '.' |
| 3 | Literal integer | 42 | 1 |
| 4 | Literal decimal | 0.314159 | 1.0 |
| 5 | Identifier | counter | p |
| 6 | Basic numerical type | int, short, long, double | num |
| 7 | Function name | main | foo() |

Figure 4. Normalisation operations on source code elements. [11, 21]

Text-based approaches are differs from one another, as they based on different techniques such as fingerprints and dot plot. Ducasse et al. [21] present one of the most recent text-based approaches that based on dot plot. The scatter plot is a two dimensional chart formed by two axes of source code. In this approach the comparison unites are lines of program. The dot appears when x and y are equal which based on the calculated hash value of both lines. The diagonals in dot plots can identify the clones, as dot plot can further be used to display the information of code clones. String-based dynamic pattern matching is applied by Ducasse et al.

on dot plot to match the lines of code. Another similar approach SDD is identified by Lee and Jeong [22], applies an n-neighbor technique to detect type-3 clones.

As the approach of Ducasse et al is not sure about recognizing the type-3 clones Wettel and Marinescu [23] provide an extension to this approach to find nearmiss clones using dot plots. They use the algorithm that relates the neighboring lines to detect some forms Type-3 clones. The approach of Marcus and Maletic [24] applies latent semantic indexing (LSI) to source code to identify and retrieve the two similar code portions, based on the similar used comments and identifiers of the retrieved code fragments. This approach limits to identify abstract data types (ADTs) which are high level concept clones.

However, Johnson is a key person who applies the technique of fingerprints in his approach on substrings of the source code [25, 26]. Initially, he uses the hash technique to hash certain portion of code a fixed number of lines (the window). Then they classify the sequences of lines which have the same hash value as clones, using the known sliding window techniques as well as the incremental hash function. The sliding window technique helps to recognise code clones of variant lengths, as it is frequently applied. Also, Manber [27] uses in his approach the technique of fingerprints to detect similar source files, based on the sub sequences of the main keywords.

### 3.3.2 METRICS-BASED TECHNIQUE

In Metrics-based approaches, several software metrics are gathered for clone fragments to derive its measurement in order to be compared instead of comparing the actual portions directly. These metrics values are related to variants scopes such as a package, a class or a method and then these values are compared to detect code clones over these blocks. The source files are normally parsed to Abstract Syntax Trees representation as a pre-process to calculate the software metrics.

There are several software metrics tools which can be used for code measurements. [28] Lincke et al. have made selection set of software metrics tools according to analyzable languages, metrics calculated, and availability/license type. They found that the majority of metrics tools available can derive metrics for many programming languages such as Java programs, UML and C/C++. They state that about half of the tools are rather simple "code counting tools" which calculates Lines of Code (LOC) metric. However, they consider the other half as more sophisticated software metrics such as CBO (Coupling Between Object classes). The following are some of the finally selected software metrics tools by Lincke et al. [28]:

- **OOMeter** this software metrics tool accepts Java / C# source code and UML models in XMI and calculates various metrics, developed by Alghamdi et al [29].
- **Eclipse Metrics Plug-in 1.3.6** this is an open source metrics calculation and dependency analyzer plugin for the Eclipse IDE. It calculates several metrics and catches cycles in package and type dependencies, developed by Frank Sauer.
- **CCCC** this is an open source command-line tool. It analyses C++ and Java code, proposed by Chidamber & Kemerer and Henry & Kafura.
- **Understand for Java** this is a metrics tool for Java source code.
- **Dependency Finder** this is an open source tool for analyzing compiled Java code. It extracts dependency graphs and mines them for useful information.
- **Semmle** is an Eclipse plug-in which allows searching for bugs and measure code metrics by providing and SQL querying languages for object oriented code.
- **Analyst4j** this is an Eclipse IDE plug-in which allows several features such as search, metrics and report generation for Java programs.

- **Eclipse Metrics Plug-in 3.4** this is an open Source tool which calculates various metrics during, developed by Lance Walton.

Lincke et al [28] also considered some software metrics derived by aforementioned tools and they base their selected metrics on the class unit, as they argue that this unit is the natural block of object oriented software systems and most metrics have been calculated on class level. The following are some considered software metrics in their study:

- **LOC** (Lines Of Code) calculates the lines of code of a specified unit [30].
- **NOM** (Number Of Methods) calculates the methods in a class [31].
- **LCOM-CK** (Lack of Cohesion of Methods) describes the lack of cohesion between the methods of a class [32]. It is proposed by Chidamber & Kemerer
- **CBO** (Coupling Between Object classes) gives the number of classes to which a class is coupled [32].
- **NOC** (Number Of Children) is the number of subclasses to a certain class in its block [32].
- **RFC** (Response For a Class) reflects the number of methods which can executed in response to an object of the class [32].
- **DIT** (Depth of Inheritance Tree) represents the maximum inheritance path from the class to the main root class [32].
- **WMC** (Weighted Methods per Class) it is the total of weights for the methods of a class [32]. However, using Cyclomatic Complexity software metric method weight can be achieved [33].
- **LCOM-HS** (Lack of Cohesion of Methods, proposed by Henderson-Sellers) describes the lack of cohesion between the methods of a class [31].

The following table shows the software metrics which calculated by metrics tools where 'x' indicates that a certain metric can be calculated by a certain metric tool.

| Tools Name | CBO | DIT | LCOM-CK | LCOM-HS | NOC | NOM | RFC | TCC | WMC |
|---|---|---|---|---|---|---|---|---|---|
| Analyst4j | x | x | x | | x | x | x | | x |
| CCCC | x | x | | | x | x | | | |
| Chidamber & Kemmerers Java Metrics | x | x | x | | x | x | x | | |
| Dependency Finder | | x | | | x | x | | | |
| Eclipse Metrics Plugin 1.3.6 | | x | | x | x | x | | | x |
| Eclipse Metrics 3.4 | | | x | x | | | | | x |
| OOMeter | x | x | x | | x | | | x | |
| Semmle | | x | x | x | x | x | x | | |
| Understand for Java | x | x | x | | x | x | | | |
| VizzAnalyzer | x | x | x | | x | x | x | x | x |

Figure 5. Tools and calculated metrics.[28]

Davey et al. [34] apply neural networks algorithm on code fragments (begin end block) and considered a certain features of specified code blocks. Their approach identifies the exact, parameterized, and near-miss clones. Balazinska et al. [35] have proposed their tool SMC (similar methods classifier) in which dynamic matching and metrics craterisation are combined. A similar approach defined by Kontogiannis et al. [36] which based on two different ways to find code clones. The first one uses direct comparison of the metrics values of the specified clone granularity. The second compares the specified blocks (code fragments), a statement-by-statement basis using uses a dynamic programming (DP) technique, as the small distance is considered as clone that caused by cut-and-paste habits. Both Patenaude et al. [37] and Mayrand et al. [12] take into account several metrics which consider some features such as names and control flow of functions to match functions with same or close metrics values as code clones. However, Calefato et al. [38] have showed that Metrics-based approaches have been applied in web documents to detect redundant web pages and clones.

### 3.3.3 TOKEN-BASED TECHNIQUE

Token approaches (lexical approaches) transform/parsed/lexed the source code into a sequence of tokens using compiler-style lexical analysis. These tokens are scanned to detect duplicated subsequences of tokens, as a result the matched tokens from the original code fragments are retrieved as clones. This technique is much better than the textual approach in term of detecting the minor code changes such as formatting and spacing.

Baker's tool Dup [39, 40] is one of the best tools that represent this approach. She used a lexical analyser to chop the line of the program to sequence of tokens. However, there are two types of tokens appear in this stage respectively, parameter tokens (identifiers and literals) and non-parameter tokens. The parameter tokens are encoded based on their occurrence in the line (position index) which helps in detecting type-2 clones whereas in the non-parameter tokens a hashing function is used. Suffix tree is used to present the prefixes of sequence of symbols. Common prefix means that tree suffixes share the same set of edges, which can be considered a clone.

CCFinder [7] of Kamiya et al.is another recent and Efficient token-based clone detection in this approach. Each line of the program is chopped to tokens using a lexer. The whole tokens of a certain source file are then chained as a single sequence. Then, the token sequence is transformed (based on the specified transformation rules such as add, change or delete tokens). Then, special tokens are used to be replaced by the identifiers (with considering types and names) across the source file, making code portions with different variable names clone pairs. The similar sub-sequences are searched among the transformed token sequence using suffix-tree based sub-string matching algorithm to be returned as clone pairs/clone classes. Finally, a suitable mapping is applied between the already obtained information of clone pair/ clone class the token-sequences and the clone pair/ clone class information of the original source code. Several tools are based on the CCFinder, such as RTF [41], which enhances the process of detecting clones by allowing the user to tailor tokenization; by using a more memory-efficient suffix-array in place of suffix trees. Also, Gemini [42], which uses scatter plots to display near-miss clones. Another pioneered clone detection technique is CP-Miner [18, 43], in which a frequent subsequence data mining technique [44] is used to find similar sequences of the original tokenised statements.

Some of the aforementioned techniques are used to detect plagiarism. SIM [45] is one of the plagiarism detection tools, which uses the dynamic programming string alignment technique to compare token sequences. Also, JPlag [46] and Winnowing [47] are examples of plagiarism detection tools which based on token based techniques.

### 3.3.4 TREE-BASED TECHNIQUE

Tree-matching approaches detect code clones by matching the similar sub-trees which obtained by parsing the source code (an abstract syntax tree). Several techniques of tree matching are used to search the related source code of the corresponding subtrees which are returned as clones pairs. Compare to token-based approach the AST is more sophisticated in order of detecting code clones as it is based on the structure of the program rather than variable names, literal values. CloneDr [16] which is lunched by Baxter et al. is one of the best tools uses AST.

The subtrees which are obtained by parsing the trees of the program; are hashed into buckets which compares the contained subtrees to each other using a tolerant tree matching.

Evans and Fraser [48] have proposed an approach in which structural abstraction of a program is achieved to handle exact and near-miss clones with gaps. Yang [49] has proposed an approach which can effectively manage the syntactic differences between the compared subtrees using dynamic programming approach. Wahler et al. [50] have converted the AST to XML to find exact and parameterized clones at abstract level. Then they have applied a data mining technique to find code clones.

However, there are several recent approaches facilitate the comparison process of subtree by applying alternative simple tree representations rather than considering the full subtree. Falke,et al. [51], who have serialized the AST subtrees as AST node sequences for which a suffix tree is then constructed which allows to detect the syntactic clones faster (at the speed of token-based techniques). Tairas and Gray [52] have proposed an approach which based on suffix trees, based on Microsoft's new Phoenix framework. Jiang et al. [53] in their tool Deckard have proposed a novel approach to detect similar trees, in which vectors are computed to approximate the structure of ASTs in a Euclidean space. Then Locality sensitive hashing (LSH) is applied to aggregate the similar vectors the using Euclidean distance metric and then detects the related code clones.

### 3.3.5 PDG-BASED TECHNIQUE

Program Dependency Graph (PDG)-based approaches [54, 55, 56] are semantically considering the source code details due to the high abstraction representation of the source code. Expressions and statements are presented by the nodes of the graph whereas control and data dependencies are represented by the edges. PDG provides more precise information than the syntactic approaches. As the PDG holds crucial information of a program such as control flow and data flow, thus it can facilitates matching the corresponding similar subgraphs (code clones) easily in a semantic way using matching algorithm. The approaches of the PDG are reliable and robust of code management. However, they lack of scalability to huge systems.

Komondoor and Horwitz [55, 57] have proposed one of the leading approaches in this technique known as PDG-DUP which uses program slicing to detect isomorphic PDG subgraphs [58]. They have also applied a sophisticated approach which aggregates the detected code clones with keeping the semantics of the source code [59, 60]. This has been used to support software refactoring by automates the procedure extraction. Furthermore, Gallagher and Lucas [61] have conducted a slicing based clone analysis experiment by argue the raised question "Are Decomposition Slices Clones?" going through the all variables in the whole system by computing program slices. They have showed the pros and cons to the raised arguments. Chen at al. [62] proposes an approach for code compaction, which considers syntactic structure and data flow. The technique has several features in embedded systems. Liu at al. [56] announce their tool which helps in plagiarism detection purposes. Krinke [54] also proposes an iterative PDG-base technique (k-length patch matching) to finding maximal similar subgraphs.

## 4. EXTENDED SMP ALGORITHM FOR CLONE DETECTION

SMP has solved several similar optimisation issues in different fields such as matching jobs to the most suitable jobseekers. Since the original SMP algorithm allows only the candidates of the first set (Men) to propose to their first choices, this research devotes to increase the fairness of SMP by allowing the candidates of the second set (Women) to make their own choices i.e. proposes to the best of their choices of the opposite set. The proposed approach considers a dual multi allocation technique that allows the candidates of both first and second set to enter the competition and propose again for a certain times to their preferences. So, each candidate of the first set may have more than one matched participants of the second set and vice versa. This adaption has enhanced the precision of the matching process; it is illustrated in Figure 6 below. In the main SMP algorithm the desire is not controlled by the similarity, thus the assigned candidates are not meant that they are similar to each other. However, in clone detection the concept of similarity is essential. Therefore, aforementioned extension of the current state of SMP is necessary to be effectively applied in such applications. A novel matching scheme is needed to achieve smart interaction between the code fragments of the matched source files. This widens the spot to detecting every possible clone.

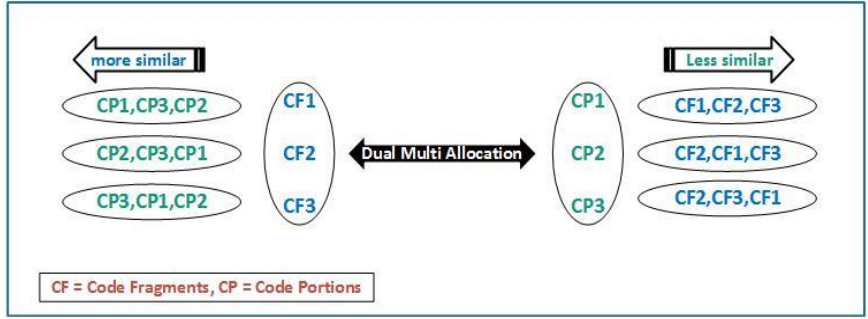

Figure 6. Dual Multi Allocation.

Practically, this process gives more than one stable matched pairs; respectively Hospital-Oriented-man and Hospital-Oriented-woman. Thus, we enclose a novel way of assigning the related code portions by adding a choosy strategy. This strategy helps to choose the pairs which form similar code clones to a certain threshold.

### 4.1. DUAL MULTI ALLOCATION ALGORITHM

Dual Multi Allocation results in several stable matching pairs with dissimilar allocated candidates based on love's degree, which can be controlled to reach a certain level of desires. However, the matching process can be fixed as default to retrieve candidates of the highest rank love's degree factor. The algorithm of Dual Multi Allocation consists of two phases followed by the Choosy Strategy as following: Phase 1 Hospital-Oriented-Man algorithm; Hospital-Oriented-Woman algorithm; Apply Choosy Strategy.

```
Algorithm 3 Dual Multi Allocation algorithm
1: assign each person to be free
2: while (some man m is unallocated ) and (m's list
   contains a woman w not allocated to m) do
3: begin
4: w:= first woman on m's list;
5: if w is already allocated, say to m', then
6: break the allocation of w to m';
7: assign w to m;
8: for each successor m' of m on w's list do
9: remove m' and w from each other's lists
10: end;
11: assign M1 to the Hospital-oriented-man pair or set of
    pairs;
12: assign each person to be free
13: while (some woman w is unallocated ) and (w's list
    contains a man m not allocated to w) do
14: begin
15: m:= first man on w's list;
16: if m is already allocated, say to w', then
17: break the allocation of m to w';
18: assign m to w;
19: for each successor w' of w on m's list do
20: remove w' and m from each other's lists
21: end;
22: assign M2 to the Hospital-oriented-woman pair or set
    of pairs;
23: apply Choosy Strategy on M1 and M2
24: end;
```

## 4.2. CHOOSY STRATEGY

In the current state of the SMP algorithms, there is no available mechanism to select the most optimal pair. Therefore, a new competitive strategy (choosy strategy) has been defined to support the newly introduced extension in choosing the optimal pair.

The choosy strategy formed out of two main factors, respectively, love's degree and contrast's degree. Love's degree reflects the degree of love from the view of both involved candidates (code fragment). To converge these views, the love's degree is defined as the average of the degrees of love for both of participated (in the same pair) candidates. The contrast's degree reflects the difference between the actual loves' degrees of the involved candidates. Thus, the most preferable pair is that with small difference in its contrast's degree. This factor helps when two different pairs has the same love's degree. Also, when more than one candidate has the same love's degree with a certain candidate, then the right candidate will be chosen. Figure 7 depicts the choosy strategy scheme.

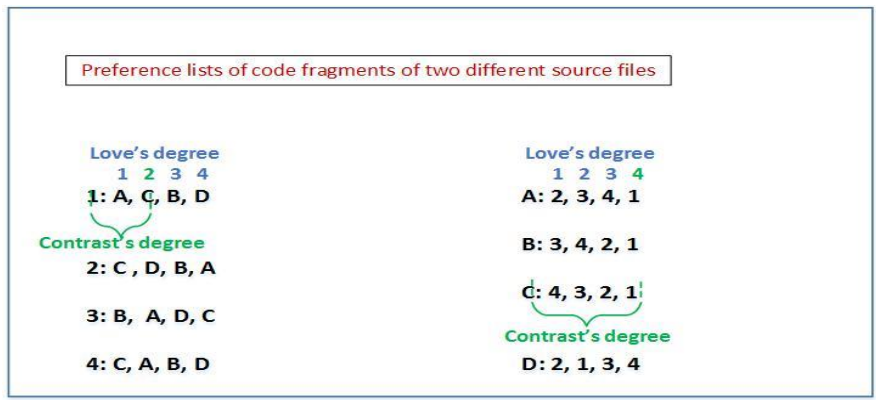

Figure 7. Choosy Strategy Scheme

## 4.3. SMP-BASED CLONE DETECTION

To apply the SMP algorithm in clone detection, it needs first to build the preference lists of both code fragments. This can be achieved using predefined metrics to specify the most similar related participants (code clone). Each code portion needs to strictly order the code fragments based on the similarity and vice versa. The traditional SMP algorithm performs a single assignment (one-to-one) for the involved candidates, which does not help especially in the case of allocating more than one code portion (method etc.) to the related code fragments of other source file. Multi Dual Allocation algorithm has been proposed to fulfil this requirement which widely needed in such fields. Figure 8 shows a small example of code clones (method-based).

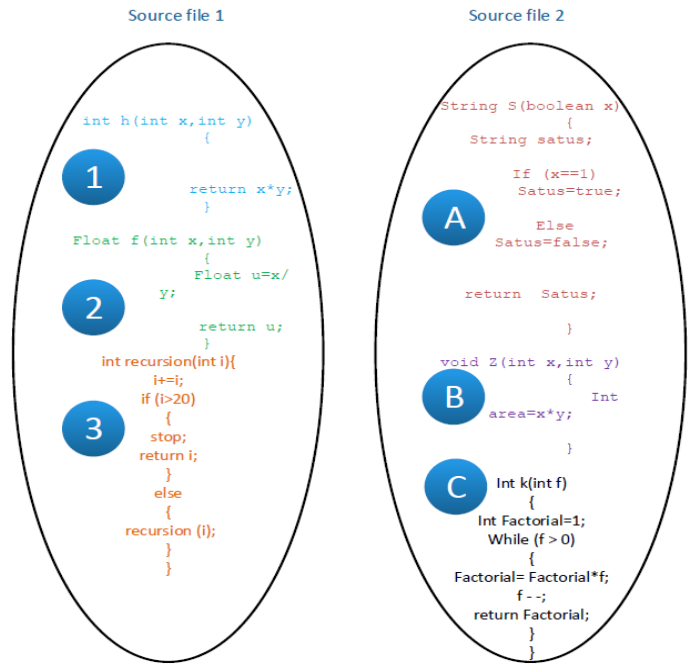

Figure 8. General Example of clones (method-based).

The metrics for fixed granularity are calculated by using java plug-in with eclipse 1.3.3 (metrics 1.3.6). Figure 9 shows some of these metrics.

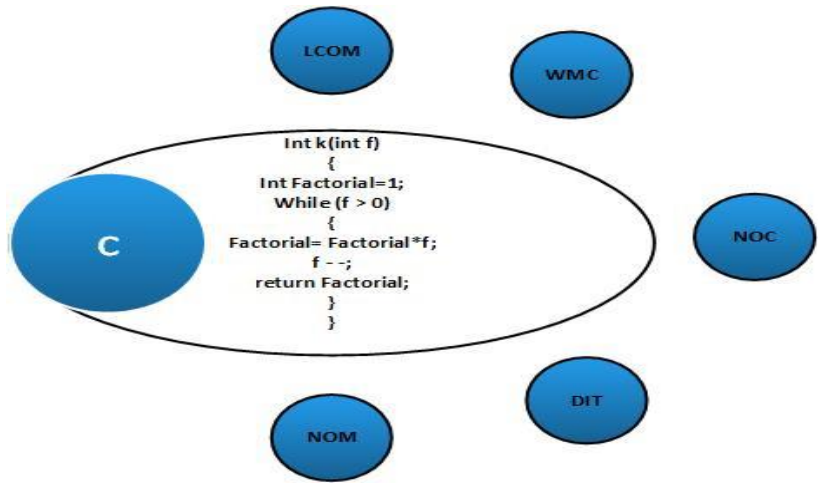

Figure 9. General view of metrics (method-based).

Table 1. Coupling Metrics

| Abbreviations | Description |
|---|---|
| PROM | Number of protected methods |
| PUBM | Number of public methods |
| PRIM | Number of private methods |
| MCIN | Number of calls to a method |
| MCOUT | Number of calls from a method |

Table 2. Method Metrics

| Abbreviations | Description |
|---|---|
| LOC | Number lines of code |
| Nbp | Number of parameters |
| Nbv | Number of variables declared in the |
| Mca | Afferent coupling at method level |
| Mce | Efferent coupling at method level |
| CC | McCabe's Cyclomatic Complexity |
| NBD | Nested Block Depth |

The extended SMP algorithm on clone detection can be applied with two main phases.

**Phase1**, building the preference list of each code fragment of the first source file from the second source file's code portions, recording the most desired block and so on, repeating this process from second to first source files.

**Phase2**, applying the adapted SMP algorithm based on the given metrics values.

Figure 10 shows an assigned code fragment of the first source file to the most suitable (similar) code portions using the adapted SMP algorithm based on the metrics values.

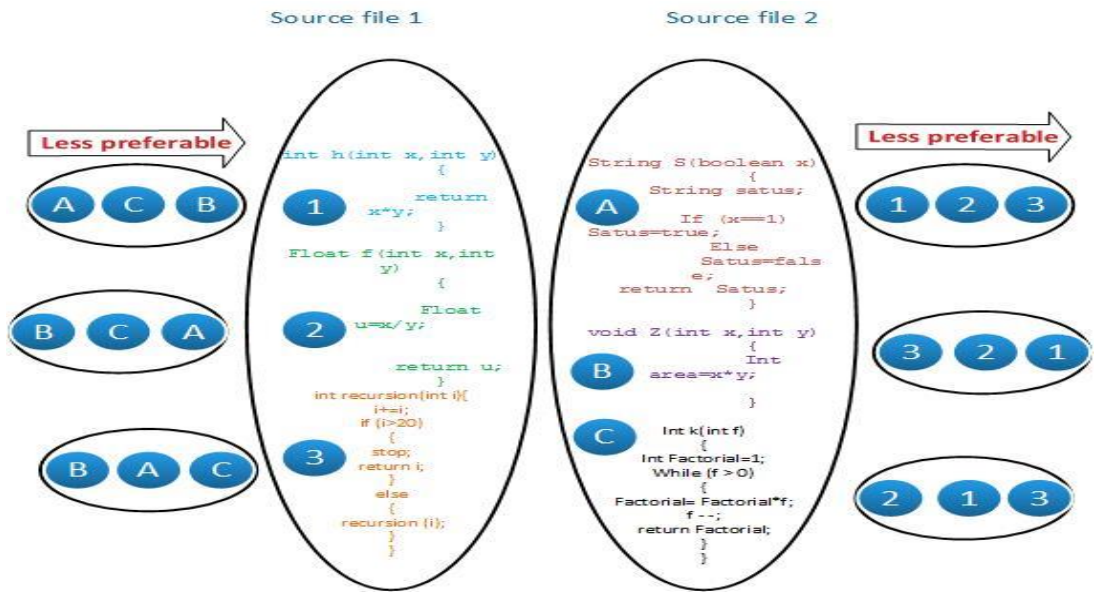

Figure 10. Example of clone detection SMP-based.

Table 3. Metrics of Source file 1

| Method | Metrics | | | | | | |
|---|---|---|---|---|---|---|---|
|  | LOC | Nbp | Nbv | Mca | Mce | CC | NBD |
| A | 6 | 1 | 1 | 0 | 0 | 2 | 1 |
| B | 4 | 2 | 1 | 0 | 0 | 1 | 1 |
| C | 6 | 1 | 1 | 0 | 0 | 2 | 2 |

Table 4. Metrics of Source file 2

| Method | Metrics | | | | | | |
|---|---|---|---|---|---|---|---|
| | LOC | Nbp | Nbv | Mca | Mce | CC | NBD |
| 1 | 1 | 2 | 0 | 0 | 0 | 1 | 1 |
| 2 | 1 | 2 | 1 | 0 | 0 | 1 | 1 |
| 3 | 10 | 1 | 0 | 0 | 0 | 2 | 2 |

The previous two tables show the measured values of the specified code blocks (methods) which act as candidates. This is a key step to build a preference list of each candidate in order to apply the proposed algorithm.

Another bigger example (a Java code of a job search system) has taken place over a medium size of source files with around 460 methods as appears in the following tables. However, the way of calculating metrics is reflecting the priority of wanted aspects of each block. This specified merged metrics are justified to accomplish the purpose of detecting as many as possible of the code clones, achieving high recall. Moreover, the precision of the retrieved code fragments needs to be considered, avoiding both false-negative and false-positive.

Table 5. Job search system (as a sample)

| #Files | Size | #Methods | #LOC | Language |
|---|---|---|---|---|
| 73 | 260Kb | 459 | 6085 | Java |

Table 5 shows the details of the job search system, which has been used to extract software clones using our approach (SMP-based). Table 6 shows some calculated metrics in the job search system.

Table 6. Snapshot of some metrics in the job search system

| Method | Metrics | | | | |
|---|---|---|---|---|---|
| | LOC | Nbp | Nbv | CC | NBD |
| AdminPage | 118 | 0 | 2 | 3 | 3 |
| ManageType | 61 | 0 | 6 | 3 | 4 |
| ChangePass | 40 | 1 | 2 | 2 | 3 |
| NewJob | 49 | 1 | 11 | 2 | 3 |
| JobSearch | 122 | 2 | 6 | 5 | 4 |
| UpdateAdmin | 48 | 1 | 6 | 2 | 2 |

The experiment shows that the SMP-based approach can precisely identify a type-3 of clone which is a copy with further modifications more than syntactic changes. The main features of SMP-based approach are competitively outweigh some already exists approaches in several facets such as increasing the spot of the detection process trying to detect every possible clone smell. However, this approach is lack to two related features respectively time complexity and speed as the used algorithm is executed in a quadratic time, which impacts the overall scalability. The clone granularity has been set as a method-based blocks in which every method acts as a candidate (possible software clone).

**4.4. DISCUSSION**

A remarkable efficiency of the proposed approach have been evaluated by carrying out a case study on two medium size source files, each file has more than 100 specified blocks. Also, a set of metrics are predefined, which help each candidate to build up its own preference list in order to apply the SMP algorithm. We observing some appointed features for the extended algorithm (e.g. performance) and the status of the detected clones (e.g. accuracy). This means that we are now able to develop match making code fragments that not only decide on the basis of the candidates' preferences of the first source file, but are actually trying to, within the current set of code fragments of both source files, to optimise the pairings from both perspectives fairly. Also, allowing the many-to-many relationship has increased the range of clones (high recall, high precision) that are undetectable with most of previous clone detection approaches. The time complexity is the same as the original SMP (polynomial time).

**5. CONCLUSION**

Stable marriage problem are well-known common matching algorithms. It has been used in many applications, for instance, assigning medical schools graduates students to the most suitable hospitals. The paper presented a newly crucial extension of SMP, which effectively touches a wide range of software engineering fields such as clone detection. The main contribution in this paper is the choosy strategy, which compromises between the preferences of the code fragments of two matched source files in clone detection process and helps to increase the quality of retrieved code clones through considering the desire of the matched candidates, which results in the increased satisfaction of the candidates in each pair. However, the proposed scheme has some limitations in terms of its complexity and would require longer time to reach the highly required stability.

**Authors**

**Hosam Al Hakami** received his B.Sc. degree in Computer Science from King Abdulaziz University, Saudi Arabia. He received his MSc degree in Internet Software Systems from Birmingham University, Birmingham, UK. He is studying now towards his PhD degree at the Faculty of Technology in De Montfort University, Leicester, UK.

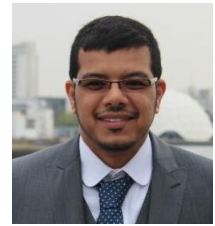

**Dr.Feng Chen** was awarded his BSc, Mphil and PhD at Nankai University, Dalian University of Technology and De Montfort University in 1991, 1994 and 2007. As research outputs, he has published over 30 research papers in the area of software evolution and distributed computing.

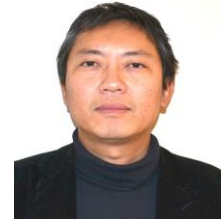

**Dr.Helge Janicke** is heading the Software Technology Research Laboratory at De Montfort University, Leicester (UK). He is leading the research theme on Computer Security and Trust. His research interests are in area of software engineering where he is primarily looking at cyber security, in particular access control and policy-based system management.

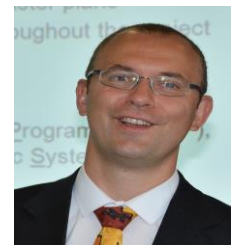